\documentclass[twocolumn,twoside,preprintnumbers,amsmath,amssymb,pacs]{revtex4}
\usepackage{epsfig,epsf}
\usepackage{graphicx}% Include figure files
\usepackage{dcolumn}% Align table columns on decimal point
\usepackage{bm}% bold math

\usepackage{fancyhdr}

\begin{document}

\title{Memory Effect and Fast Spinodal Decomposition}

\author{T. Koide} \affiliation{Instituto de F\'{\i}sica, Universidade Federal
  do Rio de Janeiro, 68528, 21941-972, Rio de Janeiro, RJ, Brazil}

\author{G. Krein} \affiliation{Instituto de F\'{\i}sica Te\'orica,
  Universidade Estadual Paulista, 01405-900 S\~ao Paulo, SP, Brazil }

\author{Rudnei O. Ramos } \affiliation{Departamento de F\'{\i}sica Te\'orica,
  Universidade do Estado do Rio de Janeiro, 20550-013 Rio de Janeiro, RJ,
  Brazil}

\received{on 21 September, 2006}

\begin{abstract}
  
  We consider the modification of the Cahn-Hilliard equation when a time delay
  process through a memory function is taken into account. We then study the
  process of spinodal decomposition in fast phase transitions associated with
  a conserved order parameter.  The introduced memory effect plays an
  important role to obtain a finite group velocity.  Then, we discuss the
  constraint for the parameters to satisfy causality.  The memory effect is
  seen to affect the dynamics of phase transition at short times and has the
  effect of delaying, in a significant way, the process of rapid growth of the
  order parameter that follows a quench into the spinodal region.
  
%  PACS numbers: 98.80.Cq; 05.70.Fh; 25.75.-q
  
\end{abstract}

\maketitle

\thispagestyle{fancy} \setcounter{page}{601}

\section{Background}

Diffusion is a typical relaxation process and appears in various fields of
physics: thermal diffusion processes, spin diffusion processes, Brownian
motions and so on.  It is empirically known that the dynamics of these
processes is approximately given by the diffusion equation.  Although the
diffusion equation has broad applicability, there exist the applicability
limitations.  First, the diffusion equation does not obey
causality\cite{ref:1}.  Let us consider the following telegraph equation,
\begin{eqnarray}
\tau \frac{d^2}{dt^2}m(x,t) + \frac{d}{dt}n(x,t) - D\nabla^2 n(x,t) = 0.
\end{eqnarray}
Then, the propagation speed is given by $v = \sqrt{D/\tau}$.  Thus, the
propagation speed of the diffusion equation is infinite because the telegraph
equation is reduced to the diffusion equation in the limit of vanishing
$\tau$.  Second, the diffusion equation does not satisfy exact relations, for
instance, the Kramers-Kronig relation and the f-sum rule
\cite{ref:Kada-M,ref:Koide}.  By solving the Heisenberg equation of motion,
the exact Laplace-Fourier transform of the time-evolution of a conserved
number density is given by
\begin{eqnarray}
\delta n^{LF}(k,z) = \frac{i}{z}C(k) + \frac{i}{z^3}\frac{k^2}{m}\langle n(0) 
\rangle_{eq} + {\cal O}(1/z^4),
\end{eqnarray}
where $\delta$ means the fluctuations from the equilibrium value and $C(k)$
represents the Fourier transform of the correlation function of the number
density.  Because of the Kramers-Kronig relation and the f-sum rule, the term
proportional to $1/z^2$ disappears and the coefficient of the second term is
proportional to the equilibrium expectation value of the total number $\langle
n(0) \rangle_{eq}$.  These are not satisfied if the coarse-grained dynamics of
the number density is assumed to be given by the diffusion equation
\cite{ref:Kada-M,ref:Koide}.

It is known that the problem of causality and the sum rules can be solved by
using the telegraph equation instead of the diffusion equation.  Interestingly
enough, it has been shown recently that the coarse-grained equation derived by
employing systematic coarse-grainings from the Heisenberg equation of motion
is not the diffusion equation but the telegraph equation
\cite{ref:Koide,ref:footnote1}.

The discussion so far is applicable to conserved quantities because the
diffusion equation is a coarse-grained equation of conserved quantities.  On
the other hand, the corresponding equation for a non-conserved quantity is the
time-dependent Ginzburg-Landau (TDGL) equation in the sense that the TDGL
equation is a overdamping equation.  The microscopic calculation
\cite{ref:Koide-M} again shows that the relaxation phenomenon is accompanied
by oscillation and cannot be described by the TDGL equation as is the case
with conserved quantities.  As a matter of fact, the equation has a similar
form to the telegraph equation.

The effect discussed above is, in particular, important for fast processes
where the time scale of memory is not short enough, because the telegraph
equation can be obtained by introducing the memory effect to the diffusion
equation.  Such fast phase transitions are expected to have happened in early
universe and most certainly also characterize the phase transitions expected
to occur in the highly excited matter created in relativistic heavy-ion
collisions.  In the early universe, such situations may have happened when the
typical microscopic time scales for relaxation, given by the inverse of the
decay width associated with particle dynamics, is larger than the Hubble time.
This is a situation likely to be expected when describing GUT phase
transitions or even the inflationary dynamics \cite{ref:Rudnei}.  In
relativistic heavy-ion collisions, one expected to learn about the QCD phase
transition.  For instance, from the hydrodynamic analysis of the freeze-out
temperature in the most central Au-Au collisions at 130 A GeV, the typical
reaction time is given by around 10-20 fm \cite{ref:Hama}.  However, the
characteristic time scale of the memory function in the Langevin equation
which describes the dynamics of the order parameter of the chiral phase
transition is predicted to be about 1 fm, which is not enough short to be
ignored \cite{ref:Koide-M}.

\section{Memory effect in spinodal decomposition}

So far, we have discussed the difficulties with causality and memory effects,
and sum rules associated with linear processes. Next, we consider such
problems associated with a nonlinear process, known as spinodal
decomposition(SD) \cite{ref:KGR}.

We consider the general Ginzburg-Landau (GL) Free energy,
\begin{eqnarray}
F(\phi) = \int d^3{\bf x}\left[
\frac{a}{2}(\nabla \phi)^2 - \frac{b}{2}\phi^2 + \frac{c}{4}\phi^4
\right],
\end{eqnarray}
where $\phi$ is a conserved order parameter.  To describe the phase
transition, the parameter $b$ is proportional to $T_c -T$ with $T_c$ being the
critical temperature of the associated second order phase transition.  The
dynamics of conserved order parameters can be described by the Cahn-Hilliard
(CH) equation.  In the ordinary CH equation, the irreversible current induced
by the GL free energy is given by
\begin{eqnarray}
{\bf J}(x,t) = -\Gamma \nabla \frac{\delta F(\phi)}{\delta \phi},
\end{eqnarray}
where $\Gamma$ denotes a kind of Onsager coefficient.  To take the memory
effect into account, the current is generalized as follows,
\begin{eqnarray}
{\bf J}(x,t) = -\int^{t}_{0}ds d^3 {\bf x}' 
{\cal M}({\bf x-x'},t-s)\nabla_{\bf x'} 
\frac{\delta F(\phi)}{\delta \phi({\bf x}',s)}. \label{GFick}
\end{eqnarray}
For simplicity, we assume the following memory function,
\begin{eqnarray}
{\cal M}(x,t) = \frac{\Gamma}{\gamma} e^{-t/\gamma}\delta^{(3)}(x).
\end{eqnarray}
Here, the typical memory time is characterized by $\gamma$.  Substituting the
current to the equation of continuity, we obtain the modified CH equation,
\begin{eqnarray}
\gamma \frac{\partial^2}{\partial t^2}\phi(x,t) 
+ \frac{\partial}{\partial t}\phi(x,t)
= \Gamma \nabla^2 \frac{\delta F(\phi)}{\delta \phi}. \label{eqn:causalCH}
\end{eqnarray}

\begin{figure}\leavevmode
\begin{center}
  \epsfxsize=7.5cm \epsfbox{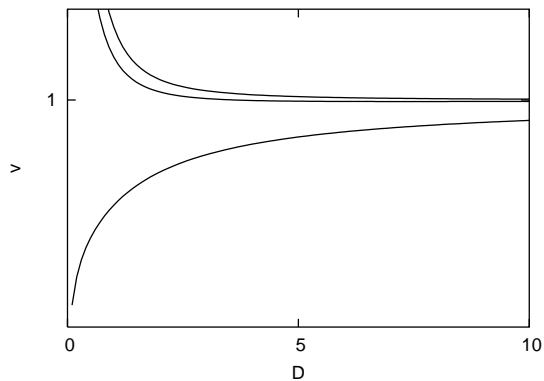}
\caption{
  The group velocity (\ref{eqn:GV}) as function of $D$ for a fixed value
  $\xi=1$.  The three lines are corresponding to $\gamma = D$, $D+0.1$ and
  $D+1$, respectively.}
\label{velocity}
\end{center}
\end{figure}

At the late stage of the spinodal decomposition, the behavior of the process
is described by the linearized equation around equilibrium order parameter
$\phi_0 = \sqrt{b/a}$,
\begin{eqnarray}
\gamma \frac{\partial^2}{\partial t^2}\tilde{\phi}_c(k,t)
+\frac{\partial}{\partial t}\tilde{\phi}_c(k,t) = 
-\Gamma k^2 (ak^2 + 2b)\tilde{\phi}_c (k,t),
\end{eqnarray}
whose solution is
$\tilde{\phi}_c(k,t) = A_k \exp(\lambda_+' t) + B_k \exp(\lambda_-' t)$,
where $A_k$ and $B_k$ are arbitrary constants and
\begin{eqnarray}
\lambda_{\pm}' = \frac{-1 \pm \sqrt{1-4\gamma \Gamma 
k^2 (ak^2 + 2b)} }{2\gamma}.
\end{eqnarray}
On the other hand, the solution for the corresponding ordinary CH equation
($\gamma=0$) is 
$\tilde{\phi}_{nc}(k,t) = C_k \exp[-\Gamma k^2 (2b+a k^2)t]$.
One can easily see that there exists a critical momentum for the solution of
the modified CH equation, ${k'}^2_c = [-b+\sqrt{b^2 + a/(4\gamma\Gamma)}]/a$.
Below the critical momentum, the solution shows overdamping behavior similar
to the ordinary CH solution, while above the critical momentum, the damping is
accompanied by the oscillatory fluctuation mode.  Thus, we can consider that
the critical momentum is a kind of a ultraviolet cutoff because the higher
momentum modes have rapid oscillations and cancel in computing averages.  The
propagation speed is characterized by the group velocity.  In the modified CH
equation, this is given by
\begin{eqnarray}
v (k) = \frac{\sqrt{2D}(\xi^2 k^2 + 1)}{\sqrt{\gamma (\xi^2 k^2 + 2)}},
\end{eqnarray}
where $D=2b\Gamma$ and $\xi=\sqrt{a/b}$.  At $k=0$, the group velocity reads
$v (0) = \sqrt{D/\gamma}$.
The maximum group velocity is a monotonically increase function of $k$.
Because of the momentum cutoff discussed above, the maximum group velocity is
given by
\begin{eqnarray}
v(k'_c) = \frac{2D}{\gamma}\frac{1+\xi^2/(2\gamma D)}
{1+\sqrt{1+\xi^2/(2\gamma D) }}. 
\label{eqn:GV}
\label{eqn:maxv}
\end{eqnarray}
For causality, Eq. (\ref{eqn:maxv}) should be less than one.  This leads to
the constraint $\gamma > D$.
In other words, we can always find allowed values of parameters for which the
spinodal decomposition is causal, under the constraint.  In Fig.
\ref{velocity}, the group velocity is plotted as function of $D$ for a fixed
value $\xi=1$.  The three lines corresponds to $\gamma = D$, $D+0.1$ and
$D+1$, respectively.  One can see that the group velocity is larger than one
for any $D$ at $\gamma = D$.  At $\gamma = D + 0.1$, one can find $D$ where
the group velocity is less than one.  At $\gamma = D + 1$, the group velocity
is less than one for any $D$.  The modified CH equation satisfies causality in
this sense.

Let us see how this also applies for instance to the spinodal modes and then
use these results e.g. for the problem of searching a signal of a phase
transition in RHIC.  By considering the fastest-growing mode of the SD, which
is defined by
\begin{eqnarray}
\left. \frac{\partial}{\partial k}
\frac{-1+\lambda}{2\gamma}\right|_{k_r} = 0,
\end{eqnarray}
we derive the time scale of the fastest mode to be
\begin{eqnarray}
\tau_c(k_r) = \frac{2\gamma}{-1+\sqrt{1+\gamma D/(2\xi^2)}}.
\label{tauc}
\end{eqnarray}
When $\gamma$ is very small, the time scale is reduced to $\tau_{nc}(k_r) =
8\xi^2/D$, which agrees with the time scale of the SD without the effect of
memory~\cite{ref:gavin2}. Since the modes that give the result of
Eq.~(\ref{tauc}) are in fact the dominant spinodal modes and $\tau_{nc} <
\tau_{c}$, this reflects itself in an overall delay of the time formation of
domains, as described by the causal CH equation compared to the noncausal one,
as the phase transition proceeds. This feature is confirmed by our simulations
shown below. {}For a problem like RHIC and a possible signature of a phase
transition coming from it, this difference in time scales can be very
pronounced and lead to a striking effect that a signal, like charge
fluctuations and domain formation, can be so much delayed that possibly could
not be observed in the current experiments. {}For instance, in RHIC, the
correlation length is typically $\xi \sim 1$ fm. Using also the relation
between the parameters $D$ and $\gamma$ obtained for a quark plasma, $\gamma =
3 D$, which is consistent with our constraint condition obtained from Eq.
(\ref{eqn:GV}), and considering $D \sim 3.7 $~fm \cite{ref:gavin2,ref:gavin},
we obtain from Eq.~(\ref{tauc}) that $\tau_c\sim 6.1$ fm, which is to be
compared with the result $\tau_{nc}\sim 2.2$ fm. This represents almost a $200
\%$ difference for the time scales for the starting of the growth of
fluctuations in Eq.~(\ref{eqn:causalCH}) as compared to the ordinary CH
equation (for $\gamma=0$).

\begin{figure}\leavevmode
\begin{center}
  \epsfxsize=7.5cm \epsfbox{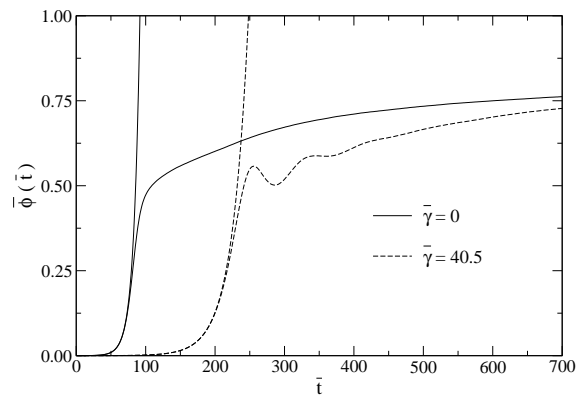}
\caption{
  Volume average of $\bar\phi(\bar t)$ as a function of dimensionless time
  $\bar t$ for the noncausal (solid) and causal (dashed) solutions. The two
  curves that exit the top of the plot are results from the linearized theory.
}
\label{fig:phi}
\end{center}
\end{figure}

We have solved Eq.~(\ref{eqn:causalCH}) numerically on a discrete spatial
square lattice using a semi-implicit scheme in time, with a fast {}Fourier
transform in the spatial coordinates~\cite{ref:KGR}.  We have checked the
stability of the results by changing lattice spacings and time steps. In
addition, for $\gamma > 1$ we have also used a leap-frog algorithm and the
results obtained with both methods agreed very well. {}For the noncausal
$\gamma = 0$ equation, we used as initial condition $\phi({\bf x},t=0)$ a
random distribution in space with zero average and amplitude $10^{-3}$. {}For
$\gamma \neq 0$, we used in addition the condition that at $t = 0$ the
first-order derivative of $\phi$ is zero. {}For the numerical work, Eq.
(\ref{eqn:causalCH}) is re-parameterized to dimensionless variables,
conveniently defined by time $\bar{t} = (8/\tau_{nc}) t= (2 b^2 \Gamma/a) t$,
space coordinates $\bar{x}_i = x_i/\xi$, field $\bar{\phi} = \sqrt{c/b} \,
\phi$ and $\bar{\gamma} = (8/\tau_{nc})\gamma$. In terms of these variables
Eq. (\ref{eqn:causalCH}) becomes function of only one parameter,
$\bar{\gamma}$. Eq. (\ref{eqn:causalCH}) was next solved for several values of
$\bar{\gamma}$. Two representatives results, for $\bar{\gamma}=0$ and
$\bar{\gamma} = 40.5$ (for the example analyzed in the previous paragraph),
are shown in {}Fig.~\ref{fig:phi}.

In {}Fig.~\ref{fig:phi} we present the time evolution of the volume average of
the order parameter $\phi$, defined as $\bar \phi(t) = \frac{1}{N^3}
\sum_{{\bf x} } \langle \bar \phi({\bf x},t)_+ \rangle$, where $N^3$ is the
total number of lattice points and the average is taken over different initial
random configurations. $\phi({\bf x},t)_+$ indicates that only the positive
values of the field are considered (i.e., a specific direction for the field
has been selected). The rapid increase of $\bar{\phi}$ reflects the phenomenon
of SD. The figure also shows the results by solving the linear equation. It
shows that it performs extremely well up to and right after the spinodal
growth of the order parameter, then justifying our previous analytical results
based on the solution of the linear equation. The effect of $\gamma$ is seen
to be more important at earlier times, consistent with the memory function
used and becomes less important after the rapid growth of the order parameter
(the spinodal explosion). It also shows the effect of a finite $\gamma$,
increasing dramatically the delay of the spinodal explosion, as predicted by
our previous analytical results. The time for reaching equilibrium is seen to
be very long, as is common with the traditional noncausal CH equation.  Also
apparent from {}Fig.~\ref{fig:phi} are the oscillations in the order parameter
for a finite $\gamma$, also predicted by our previous analysis. This is due to
the increasing importance of the second-order time derivative as compared to
the first-order one as $\gamma$ increases, i.e. as $\gamma$ increases the
dissipation term becomes less important and the equation becomes more and more
a wave-like equation. The estimated delay for the thermalization is even
larger than the recent estimation~\cite{ref:FK} for the time delay of the
relaxation of a nonconserved order parameter.

\begin{figure}\leavevmode
\begin{center}
  \epsfxsize=7.5cm \epsfbox{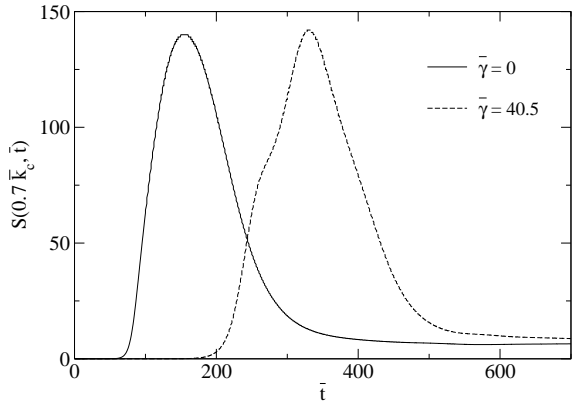}
\caption{
  Spherically averaged structure function as a function of dimensionless time
  $\bar t$ for a $\bar k = 0.7 \bar k_c$ for the noncausal (solid) and causal
  (dashed) solutions.  }
\label{fig:struct}
\end{center}
\end{figure}

We have also investigated the effect of memory for the structure function,
$S(k,t) = |\phi({\bf k},t)|^2$.  This quantity is important because it
provides information on the space-time coarsening of the domains of the
different phases. In {}Fig.~\ref{fig:struct} we present the results for the
spherically averaged value of $S$ for $k = 0.7 k_c$ (to emphasize the fast
growth of the long wavelength fluctuations with $k < k_c$). The spherical
averaging for a given $k$ was done over momenta $k_r = \sqrt{k^2_x + k^2_y +
  k^2_z}$ such that $k - 0.1 \Delta \leq k_r \leq k + 0.1 \Delta$, with
$\Delta = 2\pi/L$, where $L$ is the size of the lattice. Consistently with
Fig.~\ref{fig:phi}, this figure shows the dramatic delay for the spinodal
growth.

\section{conclusion}

We have introduced memory effects into the CH equation.  By introducing the
memory effects, we could define the group velocity and, to satisfy causality,
we were able derive the constraint for the parameters of the CH equation.
{}For a physical situation of interest for the phenomenology of RHIC, we found
that the inclusion of memory effects can delay substantially the
phase-separation process and consequently, there might not be enough time for
the system to thermalize before the breakdown of the system due to expansion.

In this paper, we simply discussed the case of memory function which has the
exponential form.  However, we can generalize the present work by considering
other types of memory functions. In particular, microscopic calculations from
nonequilibrium quantum field theory \cite{ref:Rudnei} show that memory
functions tend to exhibit relaxation as well oscillations as time goes on.
Then, we can consider the following physically motivated from those field
theory calculations,

\begin{equation}
{\cal M} ({\bf x},t) = \frac{\Gamma}{\gamma} \, e^{-t/\gamma}
\cos(\Omega t) \delta^{(3)}({\bf x}),
\label{Mkernel}
\end{equation}
where $\Omega$ is some characteristic frequency of oscillation.  Substituting
(\ref{Mkernel}) into the equation of continuity, we obtain

\begin{eqnarray}
&& \gamma^2 \frac{\partial^3}{\partial t^3} \phi ({\bf x},t) +
2 \gamma \frac{\partial^2}{\partial t^2} \phi ({\bf x},t) +
(1 + \gamma^2 \Omega^2) \frac{\partial}{\partial t}\phi ({\bf x},t) 
\nonumber \\
&& = \Gamma \, \nabla^2
\frac{\delta F(\phi)}{\delta \phi} + \gamma \Gamma 
\frac{\partial}{\partial t}\, \nabla^2 
\frac{\delta F(\phi)}{\delta \phi}. 
\label{causalCH}
\end{eqnarray}
This result and the example worked out explicitly in this work shows that the
macroscopic dynamics of conserved quantities strongly depends on the choice of
the memory function. In these cases, the use of microscopic motivated memory
functions based on the particular model under study is fundamental. An
analysis for (\ref{causalCH}), analogous to the one performed in this work for
the simplest exponential decay memory function, will be presented elsewhere.

\acknowledgments

The authors would like to thank CNPq, FAPERJ and FAPESP for the financial
support.


\begin{thebibliography}{99}
\bibitem{ref:footnote1} This microscopic calculation further makes it appear
  that the relaxation time $\tau$ contained in the telegraph equation is not
  constant and has momentum dependence.
\end{thebibliography}
\end{document}